\documentstyle[aps,multicol,graphics,times]{revtex}
\begin{document}
\draft
\title{Ubiquitous finite-size scaling features in $I$-$V$ characteristics of
various dynamic $XY$ models in two dimensions}

\author{Kateryna Medvedyeva, Beom Jun Kim, and Petter Minnhagen}
\address {Department of Theoretical Physics, Ume{\aa} University,
901 87 Ume{\aa}, Sweden}
\preprint{\today}
\maketitle

\begin{abstract}
  Two-dimensional (2D) $XY$ model subject to three different types of
dynamics, namely Monte Carlo, resistivity shunted junction (RSJ),
and relaxational dynamics, is numerically simulated. From the comparisons of
the current-voltage ($I$-$V$) characteristics, it is found that up to some 
constants $I$-$V$ curves at a given temperature are identical to each
other in a broad range of external currents. Simulations of
the Villain model and the modified 2D $XY$ model allowing stronger thermal
vortex fluctuations are also performed with RSJ type of dynamics. The
finite-size scaling suggested in Medvedyeva {\it et al.} [Phys. Rev. B (in
press)] is confirmed for all dynamic models used, implying
that this finite-size scaling behaviors in  the vicinity of the 
Kosterlitz-Thouless transition are quite robust.
\end{abstract}  

\pacs{PACS numbers: 74.76.-w, 74.25.Fy, 74.40.+k, 75.40.Gb \\
Key words: $XY$ model, current-voltage characteristics, dynamics, finite-size
scaling}

\begin{multicols}{2}

\section{Introduction} \label{sec:intro}
The phase transition between the superconducting and normal states in
many two-dimensional (2D) systems is of Kosterlitz-Thouless (KT)
type~\cite{kosterlitz}. The thermally excited vortices in a large enough
sample, interacting via a logarithmic potential, are bound in neutral pairs
below the KT transition temperature $T_{KT}$~\cite{kosterlitz,minnhagen:rev},
and as the temperature $T$ is increased across $T_{KT}$ from below
these pairs start to unbind. The KT transition has been
observed in experiments on superconducting films~\cite{kadin,repaci}, 2D
Josephson junction arrays~\cite{herbert}, and  cuprate
superconductors~\cite{gorlova}. In these experiments, the current-voltage 
($I$-$V$) characteristics have been commonly measured to detect the transition.
For small enough currents it has a power law form $V \propto I^a$ (or equivalently, $E
\propto J^a$ with the electric field $E$ and the current density $J$), where
the $I$-$V$ exponent $a$ is known to have a universal value 3 precisely at the
transition~\cite{minnhagen:rev}; For $T<T_{KT}$ one has $a > 3$, whereas $a=1$
for $T>T_{KT}$~\cite{nelson}. The dynamic critical exponent $z$ which relates
the relaxation time $\tau$ to the vortex correlation length $\xi$ via $\tau
\sim \xi^{z}$, is connected with $a$ through the relation
$a=1+z$~\cite{fisher}. Consequently, $z$ has the value 2 at $T_{KT}$.

Indeed, the values $z=2$ and $a=3$ at the KT transition have been confirmed in
many numerical simulations, e.g., the lattice Coulomb gas with Monte Carlo
dynamics~\cite{lee}, the Langevin-type molecular dynamics of Coulomb gas
particles~\cite{holmlund}, and the 2D {\it XY} model both
with the resistively shunted
junction dynamics (RSJD) and the relaxational dynamics (RD)~\cite{beom:big,melwyn}. That is why $z \approx 6$ obtained by
Pierson {\it et al.} in Ref.~\onlinecite{pierson} at the resistive transition
in many 2D systems is very intriguing. We in Ref.~\onlinecite{medv}
have re-analyzed the experimental $I$-$V$ characteristics for an ultra thin 
YBCO sample in Ref.~\onlinecite{repaci} and compared with the 2D RSJD
model. This led to the suggestion that a novel finite-size type
scaling effect of the
$I$-$V$ characteristics (which can possibly be caused not only by the actual
finite size of the sample but also by a finite perpendicular penetration depth
or a residual weak magnetic field~\cite{minnhagen:rev,medv}) is responsible for
the large value of $z \approx 6$ in Ref.~\onlinecite{pierson}, but at the same time it
was concluded that this exponent has nothing to do with the true dynamic 
critical exponent which at $T_{KT}$ has value two~\cite{medv}. 
Since similar conclusions were also reached in Ref.~\onlinecite{pierson1},
we believe that there now seems to emerge some consensus~\cite{medv,pierson1},
although the physical mechanism causing this finite-size scaling behavior
is still unclear.

 In the present paper we search for a possible reason of this finite-size scaling 
behavior.
We use the RSJD~\cite{beom:big,melwyn,simkin}, the RD 
(often referred to as time-dependent Ginzburg-Landau
dynamics)~\cite{beom:big,melwyn}, and the Monte Carlo dynamics (MCD)
simulations~\cite{lee,weber,beom} to study the $I$-$V$ characteristics of three
different models of 2D superconductors: the Villain model~\cite{villain},
the $XY$ model in its original form, and the $XY$ model modified with $p$-type of
potential (see Ref.~\onlinecite{p-pot} for details). These three models
differ from each other by the density of the thermally created vortices.
We come to the conclusion that the qualitative features of the
scaling suggested in
Ref.~\onlinecite{medv} are ubiquitous and independent of both  the vortex
density and the type of dynamics. We also demonstrate that the $I$-$V$ curves
obtained with different types of dynamics, up to constant scale factors,
coincide very well in a broad range of the external current density.
This opens a possibility to use MCD simulation, which from a simulation point of view is more efficient
than RSJD or RD, to examine long-time dynamic properties of the models.  

 The layout of the paper is as follows: In Sec.~\ref{sec:models} we recapitulate the
Hamiltonians for the generalized 2D {\it XY} model with the $p$-type of
potential (the usual {\it XY} model is recovered when $p=1$) and the Villain model,
and describe details of the dynamics used (RSJD, RD, and MCD). The results from the
usual {\it XY} model with $p=1$ subject to different dynamics (RSJD, RD, and MCD)
are presented in Sec.~\ref{sec:xy}, while the results from RSJD simulations
applied to the different types of models, i.e., the usual {\it XY} model with $p=1$, 
the {\it XY} model with $p=2$ and the Villain model, are described and
analyzed in Sec.~\ref{sec:rsj}. We summarize and make final remarks  in
Sec.~\ref{sec:conc}.

\section{Models and details of dynamics} \label{sec:models}
\subsection{Model Hamiltonians}
 The 2D $XY$ model  defined on a square lattice, where each lattice point $i$
is associated with the phase $\theta_{i}$ of the superconducting order
parameter, is often used for studies of the
KT transition. The phase variables in this model interact via the Hamiltonian
with nearest neighbor coupling, which in the absence of frustration is given by 
\begin{eqnarray}\label{eq:HXY}
H = \sum_{\langle ij\rangle}U(\phi_{ij} = \theta_{i} - \theta_{j}) , 
\end{eqnarray}
where $\langle ij\rangle$ denotes sum over nearest neighbor pairs, $\phi$ is
the angular difference between nearest neighbors, and the
interaction potential $U(\phi)$  is written as
\begin{eqnarray}\label{eq:UXY}
U(\phi) \equiv E_{J}(1-\cos\phi) ,
\end{eqnarray}
with the Josephson coupling strength $E_{J}$.

The dominant characteristic physical features  close to the KT transition are
associated with vortex pair fluctuations. One interesting aspect is then how
the density of the thermally excited vortex fluctuation effects the 
critical properties.
To study this we generalize the interaction potential by using a
parameter $p$~\cite{p-pot}:
\begin{eqnarray}\label{eq:Up}
U^{p}(\phi) \equiv 2E_{J} \left[ 1-\cos^{2p^{2}}\left({\phi\over 2}\right)\right] ,
\end{eqnarray}  
where $U^{p=1}(\phi)$  corresponds to the potential of the usual $XY$  model
[see Eq.~(\ref{eq:UXY})]. The practical point with such generalization is that
the vortex density increases with increasing $p$~\cite{p-pot}.
The variation of the parameter $p$ can also change the nature of the transition:
for $p$ exceeding  some maximum value ($p > p_{\rm max} \approx 5$) the type of the
phase transition changes from KT to the first order~\cite{p-pot,anna}. In the
present paper we choose $p=2$ which is well inside the KT transition region,
yet is large enough to ensure substantially more vortex fluctuation over a
temperature region around the phase transition in comparison with
the usual $p=1$
$XY$ model given by Eq.~(\ref{eq:UXY}). 
While the $XY$ model with the $p$-type potential with $p>1$ has more
vortices than the usual $p=1$ $XY$ model, we also study the Villain
model~\cite{villain} which has less vortex-antivortex pairs~\cite{olsson:private}.
The interaction potential $U(\phi)$ in the Villain model is given by
\begin{equation}\label{eq:UVillain}
    e^{-U(\phi)/T} \equiv \sum_{n=-\infty}^{\infty}
    \exp\left[ -\frac{E_J}{2T}(2\pi n - \phi)^{2} \right] .
\end{equation} 

 To simulate the  dynamic behaviors of these models we use several types of
dynamics: RSJD, RD, and MCD. All these dynamics should result in the same
equilibrium static behaviors if we apply them to the models with the same
interaction potential. However, dynamic properties of the systems can be
different. Of course, different types of dynamics have their own advantages and
disadvantages.  The RSJD is constructed from the elementary Josephson relations
for single Josephson junction that forms the array units, plus Kirchhoff's
current conservation condition at each lattice site~\cite{beom:big,melwyn}.
Therefore, this type of dynamics has a firm physical realization. On the other
hand,  RSJD is quite slow which leads to the limitation in the time scale one
can probe in simulations. Although the RD~\cite{beom:big,melwyn,anna} is much
easier to implement than RSJD, it does not converge much faster than RSJD and
it does not have a similar direct physical realization as RSJD. However, a
superconductor has been argued to have a RD type of dynamics rather than a
RSJD~\cite{fisher,dorsey}.
The MCD simulations~\cite{beom} are much faster than
RSJD or RD, which allows one to investigate dynamic behaviors
in much longer time scale (one can also study dynamic behaviors 
at much lower temperatures with MCD).
However, since there is no direct physical realization of the MCD in practice
the applicability  of this dynamics to a specific physical system must
then be explicitly demonstrated.  
In the following discussions on the  details of the different dynamics used,
we focus on the original 2D $XY$ model with the interaction potential
in Eq.~(\ref{eq:UXY}) since the extensions to a modified 2D $XY$
model (\ref{eq:Up}) and Villain model (\ref{eq:UVillain}) are straightforward.

\subsection{Dynamic models}
In this section we briefly review the dynamical equations of motion for RSJD,
RD, and MCD, in the presence of the fluctuating twist boundary
condition (FTBC)~\cite{beom:big,beom}.
We perform simulations of unfrustrated square $L \times L$ lattices with 
$L=6$, 8, and 10 at various temperatures to measure the voltage across the
lattice as a function of the external current. Although the system sizes
are relatively small, which is inevitable because of the low temperatures
and the small external currents used here, the FTBC has been shown
to be very efficient in reducing the artifact due to small system
sizes~\cite{beom:big,mychoi:bc}, and  reliable results can be
 established~\cite{beom:big,melwyn,beom}.

In the FTBC, the twist variable ${\bf \Delta} = (\Delta_x, \Delta_y)$
is introduced and the phase difference $\phi_{ij}$ on the bond
$(i,j)$ is changed into $\theta_{i}-\theta_{j}- {\bf r}_{ij}\cdot{\bf \Delta}$,
with the unit vector ${\bf r}_{ij}$  from site $i$ to site $j$, 
while the periodicity on $\theta_i$  is imposed: $\theta_i=\theta_{i+L\hat{\bf
x}}=\theta_{i+L\hat{\bf y}}$. The Hamiltonian of 2D $L \times L$ $XY$ model
under FTBC without external current has been introduced
in Ref.~\onlinecite{olsson}, and is written as [compare with Eq.~(\ref{eq:HXY})]
 \begin{equation} \label{eq:HFTBC}
  H=-E_{J}\sum_{\langle ij\rangle}\cos (\phi_{ij} \equiv 
\theta_i-\theta_j-{\bf r}_{ij}\cdot{\bf
\Delta}), 
\end{equation}
which later in Ref.~\onlinecite{beom} has been extended to the system
in the presence of an external current and written as 
\begin{equation} \label{eq:HFTBCJ}
H=-E_{J}\sum_{\langle ij\rangle}\cos\phi_{ij} + \frac{\hbar}{2e}L^2 J\Delta_x,
\end{equation}
where $J$ the current density in the $x$ direction.

\subsubsection{RSJD and RD} \label{subsubsec:RSJDRD}

We introduce first the RSJD equations of motion for phase variables
and twist variables, which 
are generated from the local (global) current conservation 
for the phase (twist) variables (see
Ref.~\onlinecite{beom:big} for details and discussions).
The net current $I_{ij}$ from site $i$ to site $j$ is the sum of
the supercurrent $I_{ij}^s$, the normal resistive current $I_{ij}^{n}$, 
and the thermal noise current $I_{ij}^{t}$: 
$I_{ij} = I_{ij}^{s} + I_{ij}^{n} + I_{ij}^{t}$. 
The supercurrent is given by the Josephson current-phase relation,
$I_{ij}^s = I_{c}\sin\phi_{ij}$, where $I_{c}=2eE_{J}/\hbar$ is the critical current
of the single junction. The normal resistive current is given by
$I_{ij}^{n} = V_{ij}/r$, where $V_{ij}$ is the potential difference across the 
junction, and $r$ is the shunt resistance.
Finally the thermal noise current $I_{ij}^t$ in the shunt at temperature $T$
satisfies $\langle I_{ij}^{t} \rangle = 0$ and
$  \langle I_{ij}^{t}(t)I_{kl}^{t}(0) \rangle = (2k_BT /r)\delta(t)
  (\delta_{ik}\delta_{jl} - \delta_{il}\delta_{jk})$,
where $\langle \cdot \cdot \cdot \rangle$ is thermal average, and $\delta(t)$ and
$\delta_{ij}$ are Dirac and Kronecker delta, respectively. Using the
current conservation law at each site of the lattice together with the
Josephson relation $\dot{\phi}_{ij} \equiv d\phi_{ij}/dt = 2eV_{ij}/\hbar$ 
one can derive the RSJD equations of motion for phase variables:
\begin{equation} \label{eq:rsjphase}
\dot{\theta}_i=-\sum_j G_{ij}{\sum_k}^\prime
(\sin\phi_{jk}+\eta_{jk}) , 
\end{equation}
where the primed summation is over the four nearest neighbors of $j$, $G_{ij}$
is the lattice Green function for 2D square lattice, 
and $\eta_{jk}$ is the dimensionless thermal noise current defined
by $\eta_{jk} \equiv I_{jk}^{t}/I_{c}$. The time, the current, the distance,
the energy, and the temperature are normalized in units of $\hbar/2erI_{c}$,
$I_{c}$, the lattice spacing $a$, the Josephson coupling strength $E_{J}$, and
$E_{J}/k_{B}$, respectively. 
In order to get a closed set of equations we further specify the
dynamics of the twist variable $\bf{\Delta}$ from the
condition of the global current conservation that the summation of 
the all currents through the system in each direction should vanish~\cite{beom:big}:
\begin{eqnarray} 
\dot\Delta_x & = & \frac{1}{L^2}\sum_{\langle ij\rangle_x}
\sin\phi_{ij}+\eta_{\Delta_x} - J, \label{eq:delta_x} \\
\dot\Delta_y & = & \frac{1}{L^2}\sum_{\langle ij\rangle_y}
\sin\phi_{ij}+\eta_{\Delta_y} , \label{eq:delta_y} 
\end{eqnarray}
where $\sum_{\langle  ij\rangle_x}$ denotes the summation over all nearest
neighbor links in the $x$ direction, and we apply
the external dc current with the current density $J$ in the $x$ direction.
Here, the thermal noise terms  $\eta_{\Delta_x}$ and $\eta_{\Delta_y}$ 
obey the conditions $\langle
\eta_{\Delta_x}\rangle = \langle\eta_{\Delta_y}\rangle 
=\langle\eta_{\Delta_x} \eta_{\Delta_y}\rangle =0$, and
$\langle\eta_{\Delta_x}(t)\eta_{\Delta_x}(0)\rangle=
\langle\eta_{\Delta_y}(t)\eta_{\Delta_y}(0)\rangle=(2T/L^2)\delta(t)$.

In the RD, the equations of motion for the phase variables are 
written as~\cite{beom:big,melwyn,tdgl} 
\begin{equation}\label{eq:rdphase}
\dot\theta_{i} =
- \Gamma {\partial H \over \partial \theta_i} + \eta_{i}
 = - {\sum_j}^\prime\sin\phi_{ij}+ \eta_i, 
\end{equation}
where $\Gamma$ is a dimensionless constant (we set $\Gamma = 1$ from now one),
$H$ is in Eq.~(\ref{eq:HFTBCJ}), $t$ is in units of $\hbar/\Gamma E_{J}$, 
and the thermal noise $\eta_{i}(t)$ at site $i$ satisfies 
$\langle \eta_{i}(t) \rangle = 0$ and $\langle \eta_{i}(t)\eta_{j}(0)\rangle = 
2T\delta(t)\delta_{ij}$. 
The equation of motion for the twist variables in the absence of
an external current is of the form (see
Ref.~\onlinecite{beom:big,melwyn} for more details)
\begin{eqnarray}\label{eq:rdtwist}
\dot{\bf \Delta} = -\frac{1}{L^2} {\partial H\over \partial {\bf \Delta}} 
+ \eta_{\bf \Delta}, 
\end{eqnarray}
which is the same as Eqs.~(\ref{eq:delta_x}) and (\ref{eq:delta_y}) for RSJD.
Accordingly, to some extent the RD may be viewed as a simplified version of the RSJD
where the global current conservation is kept but the local current conservation
is relaxed. 

Consequently, the equations for the phase variables are different for RSJD and RD
[Eqs.~(\ref{eq:rsjphase}) and (\ref{eq:rdphase}), respectively]
while the same equations (\ref{eq:delta_x}) and (\ref{eq:delta_y})
apply to the twist variables for both dynamics.
These coupled equations of motion are discretized in time with the
time step $\Delta t = 0.05$ and $\Delta t = 0.01$ for RSJD and RD,
respectively, and numerically integrated using the second order
Runge-Kutta-Helfand-Greenside algorithm~\cite{rg}. The voltage drop $V$ across
the system in the $x$ direction is written as $V=-L \dot\Delta_x$ (see
Ref.~\onlinecite{beom:big}) in units of $I_c r$ for RSJD and in units of
$\Gamma E_{J}/2e$  for RD, respectively. We measure the electric field 
$E =\langle V  / L\rangle_t$ to obtain $I$-$V$ characteristics, where
$\langle \cdots \rangle_t$ denotes the time average performed
over $O(10^6)$ time steps for large currents for both RSJD and RD,
and $O(5\times10^7)$ and $O(10^9)$ steps for small currents for 
RSJD and RD, respectively. 

\subsubsection{MCD} \label{subsubsec:MCD}

The technique to simulate 2D {\it XY} model with MCD is based on the Hamiltonian
(\ref{eq:HFTBCJ}) and the standard Metropolis algorithm~\cite{metropolis}. The
one MC step, which we identify as a time unit, is composed as follows~\cite{beom}:
\begin{enumerate}
\item \label{step:pickphase}
Pick one lattice site and try to rotate the phase angle at the site by
an amount randomly chosen in $[-\delta \theta, \delta \theta]$  (we call
$\delta \theta$ the trial angle range). The twist variable ${\bf \Delta}$ 
is kept constant during the update of the phase variables.
\item \label{step:dEphase}
Compute the energy difference $\Delta H$ before and after
the above try; If $\Delta H < 0$ or if $e^{-\Delta H /T}$ is greater than a random number
chosen on the interval $[0,1)$, accept the trial move.
\item \label{step:repeatphase}
Repeat steps~\ref{step:pickphase} and \ref{step:dEphase} for all the 
lattice sites to update the phase variables.
\item\label{step:picktwist}
Update the fluctuating twist variables $\Delta_{x}$ in the
similar way that $L\Delta_{x}$ is tried to rotate within the
angle range $\delta\Delta$ with $\theta_i$ and $\Delta_y$ kept unchanged.
(For convenience, we use $\delta \Delta = \delta \theta$).
\item\label{step:dEtwist}
Compute the energy difference $\Delta H$ before and
after the trial step~\ref{step:picktwist} for $\Delta_x$. 
Accept the step~\ref{step:picktwist},
if $\Delta H < 0$  otherwise accept it with probability $e^{-\Delta H/T}$
like in step~\ref{step:dEphase}.
\item \label{step:repeattwist}
Repeat steps~\ref{step:picktwist} and \ref{step:dEtwist} 
to update $\Delta_y$.
\end{enumerate}
In the MCD simulation the trial angle
$\delta \theta = \pi/6$ has been chosen since it is sufficiently small in order
to obtain the correct $I$-$V$ characteristics while it is big enough to make
MCD much faster than the other dynamic methods~\cite{beom}.
The time-averaged electric field is obtained after equilibration from
the averages over $O(10^9)$ (at large currents) to $O(10^{10})$ (at small
currents) MC steps.

\section{2D {\it XY} model subject to different types of dynamics} 
\label{sec:xy}
In this section we use three different types of dynamics, the RSJD, the RD, 
and the MCD to study dynamic behavior of the 2D {\it XY}
model with $p=1$ under the FTBC. The dynamic behavior of the system can be obtained from  the complex conductivity, the flux noise spectrum, as well as the $I$-$V$ characteristics which is commonly measured in experiments. We will focus on the $I$-$V$ characteristics in the present paper.  
As pointed out in Sec.~\ref{sec:models} the RD to some extent may be considered
as a simplified version of the RSJD. Thus from this point of view it is perhaps
not surprising that these two models (as we will see) contain  similar
features of the vortex dynamics. In Ref.~\onlinecite{mc} from the study of a
simple dynamic model of isolated magnetic particles in a uniform field it has
been shown that the actual dynamics of the model and MCD are in a good
agreement when the acceptance ratio of the Metropolis  step is low enough. This
implies that the MCD should give the same $I$-$V$ characteristics as the RSJD
after an appropriately chosen normalization of time, when the trial angle
$\delta \theta $ is sufficiently small (it was shown in Ref.~\onlinecite{beom}
that $\delta \theta=\pi/6$ is sufficiently small). We will confirm this further
in the present simulations.

\begin{figure}
\centering{\resizebox*{!}{5.5cm}{\includegraphics{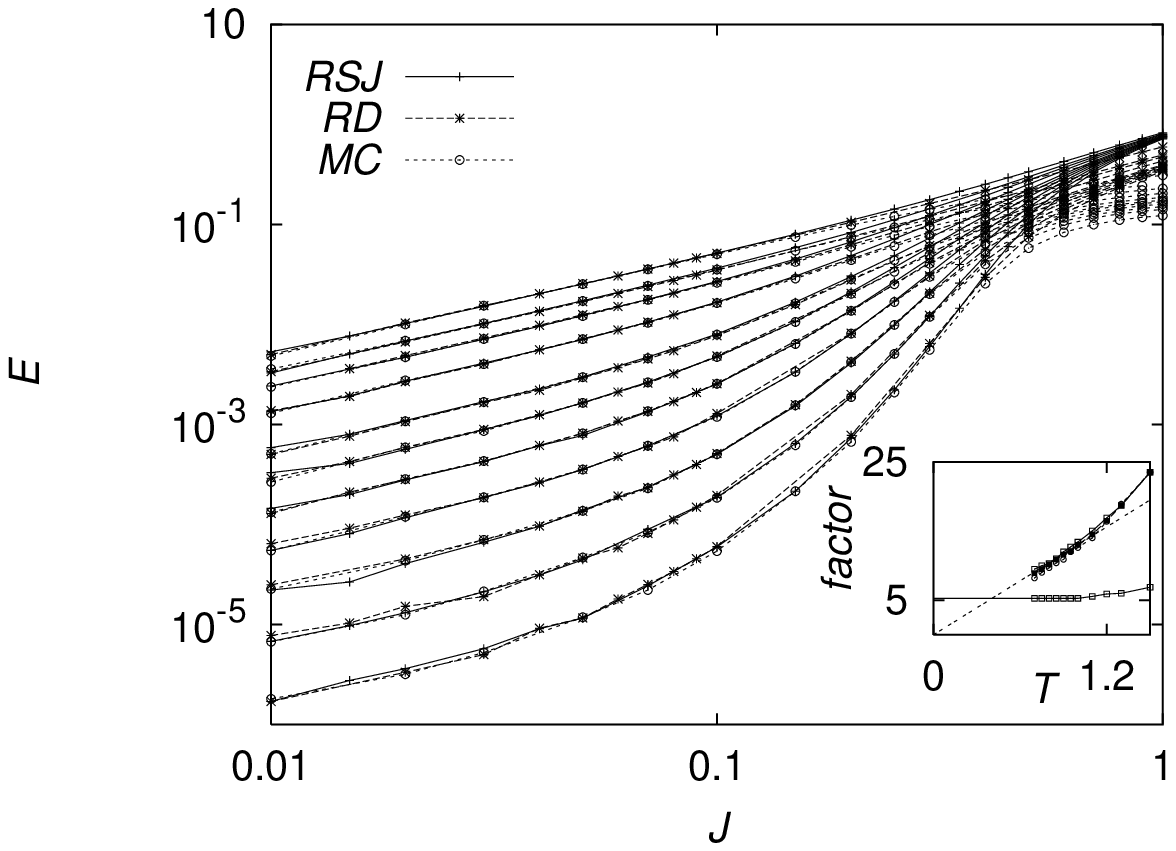}}}
\vskip 0.5cm
\caption{Comparisons of the $I$-$V$ characteristics for the 2D $XY$ model with
MC, RSJ, and RD dynamics on a square lattice with the finite size $L=8$ plotted
as $E=V/L$ against $J=I/L$ in log scales at temperatures $T=1.5$, 1.3, 1.2,
1.1, 1.0, 0.95, 0.90, 0.85, 0.80, 0.75, and 0.70 (from top to bottom).
Multiplication by a factor (presented in the inset), which in the case of MCD
depends on temperature, makes the curves coincide in a broad range of external
current density. The straight line from the origin in the inset shows that the
factor in the MCD case is a linear function of temperature for system sizes,
$L=6$, 8, 10 in accordance with expectation~\protect\cite{mc}. However, for higher $T$
there is a deviation. The horizontal full line in the inset shows that the
factor between the RSJD and RD $I$-$V$ curves is independent of $T$ (the factor
is multiplied by $10$ in the inset). }
\label{fig:IV}
\end{figure}

In Fig.~\ref{fig:IV} we compare $I$-$V$ characteristics in the form
of the electric field $E$ versus the current density $J$ 
obtained from RSJD, RD, and MCD simulations in the temperature range 
$0.70 \leq T \leq 1.50$ (the temperature interval is 0.05 if
$0.70 \leq T \leq 1.00$ and 0.10 otherwise) 
in log scales for the system size $L=8$. 
In order to make $I$-$V$ data of RSJD and RD simulations
coincide, $E$ obtained with RD ($E_{RD}$) is multiplied  by a temperature-independent 
factor represented by the horizontal line in the inset of
Fig.~\ref{fig:IV}. Since the measured time-averaged electric field is inversely
proportional to the time scale for a given dynamics, 
one can  from the ratio $E_{RSJD}/E_{RD}$ infer the correspondence between
times of RD and RSJD.
>From this comparison, we find for $0.70 \leq T \leq 1.00$
that one unit of time in RD approximately corresponds to 0.526 time unit
in RSJD, independent of the temperature. Note that the $I$-$V$ curves corresponding to
the  temperatures exceeding 1.00 are almost a straight lines. Therefore the
collapse of data between the different dynamics is trivial for $T>1.00$. Also the $I$-$V$
characteristics obtained from MCD can be made to collapse on top of the
corresponding curves for the RSJD and RD, as shown in Fig.~\ref{fig:IV}.
However, in this case the time scale factor, describing how many
RSJD time steps one MCD step corresponds to, depends on the temperature, as
shown in the inset of Fig.~\ref{fig:IV}, where this factor is shown
to be a linear function of temperature for system size $L=6$, 8, and 10 
up to $T=1.00$. This is in accordance
with the model studied in Ref.~\onlinecite{mc}, where the same 
linear behavior in terms of the temperature has been found. 
Thus for the $I$-$V$ curves we have a precise relation between 
RSJD and MCD: For example, at $T=0.80$ we get $1$ MCS = $10.2
\times$RSJD time unit. This opens a practical possibility to study some aspects
of dynamic behavior of the $XY$ model by using MCD simulation, which
is usually much more efficient than RSJD and RD. 

We have shown in this section that the RSJD, the RD, and
the MCD applied to the usual $XY$ model 
gives basically the identical $I$-$V$ characteristics up to
some constant factors. From this observation, one can conclude
that the dynamic critical behaviors of the $XY$ model inferred
from the $I$-$V$ characteristics should be identical for
all these dynamic models. 
We in next section use the RSJD to study the Villain model
and $XY$ models with $p=2$ and presume from the observation
in the present section that the conclusion
drawn in Sec.~\ref{sec:rsj} for the RSJD case should be also valid in the
other dynamics (RD and MCD).

\section{Various $XY$ models subject to RSJD} \label{sec:rsj}
 To study the critical behavior of the system in the vicinity of the transition
one can use  scaling relations. Fisher, Fisher, and Huse (FFH) in
Ref.~\onlinecite{fisher} proposed that the nonlinear $I$-$V$ characteristics
in a $D$-dimensional superconductor scales as
 \begin{equation} \label{eq:p1}
E=J\xi^{D-2-z}\chi_\pm(J\xi^{D-1}/T),
\end{equation}
where $\xi$ and $z$ are the correlation length and the dynamic critical exponent, 
respectively, and $\chi_\pm$ is the scaling function above ($+$) 
and below ($-$) the transition. Pierson {\it et al.} in
Ref.~\onlinecite{pierson} have applied a variant of this FFH scaling approach
to the $I$-$V$ data for thin ($D=2$) superconductors and superfluids and
suggested a phase transition with $z \approx 6$. In Ref.~\onlinecite{medv} 
another scaling relation has been introduced  and it has
been shown that a certain finite-size effect which is not included in the FFH
scaling may have caused the large  $z$ in spite of the fact that  the finite size
effect precludes the possibility of a real phase transition. This finite-size
scaling around and below the KT transition is given by the form (see
Ref.~\onlinecite{medv} for the details)
\begin{equation} \label{eq:ourscale}
\frac{E}{JR}=h_T \bigglb( JLg_L(T) \biggrb) , 
\end{equation}
where $R = \lim_{J \rightarrow 0}(E/J) \propto L^{-z(T)}$ is a finite-size
induced resistance without external current, $h_T(0)=1$, $h(x)\propto x^{z(T)}$
for large $x$, and $g_L(T)$ is a function of at most $T$ and $L$ such that a
finite limit function $g_\infty(T)$ exists in the large-$L$ limit. For small
values of the variable $x = JLg_L(T)$  the $T$-dependence of the scaling is
absorbed in a function  $g_{L}(T)$ for each fixed size $L$ giving rise to the
scaling form 
\begin{equation} \label{eq:scale}
\frac{E}{JR}=h\bigglb(JLg_L(T)\biggrb) .
\end{equation}

\begin{figure}
\centering{\resizebox*{!}{5.5cm}{\includegraphics{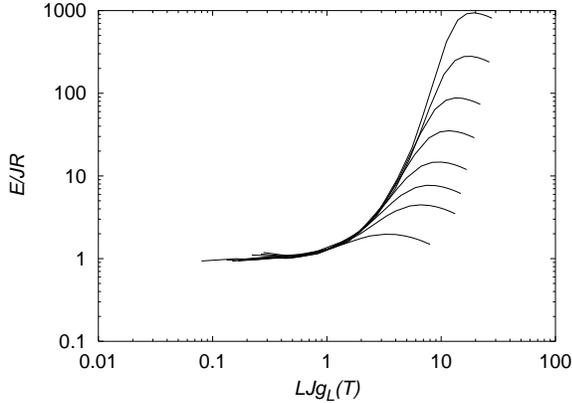}}}
\vskip 0.5cm
\caption{The finite-size scaling of the current-voltage characteristics, $E/JR$
vs  $LJg_{L}(T)$, given by Eq.~(\ref{eq:scale}). The function $g_{L}(T)$ is
determined from  data collapse for small values of $LJg_{L}(T)$. The data was
obtained from MCD and was scaled with a $T$-dependent factor so as to
correspond to RSJD and RD; system size $L=8$ and $1.10 \leq T \leq 0.70$ where
each curve corresponds to a fixed $T$.}
\label{fig:scaling}
\end{figure}

In Fig.~\ref{fig:scaling} we demonstrate the existence of the finite-size
scaling given by Eq.~(\ref{eq:scale}) for $L=8$ within the temperature
intervals $0.70 \leq T \leq 1.10$. The data are obtained for  MCD and scaled
by the appropriate factor so as to correspond to RSJD and RD (compare
Fig.~\ref{fig:IV}). Because of the correspondence between the three types of
dynamics (see Sec.~\ref{sec:xy}), this also means that the existence of the finite-size scaling given
by Eq.~(\ref{eq:scale}) is insensitive to the choice of dynamics.  

\begin{figure}
\centering{\resizebox*{!}{5.5cm}{\includegraphics{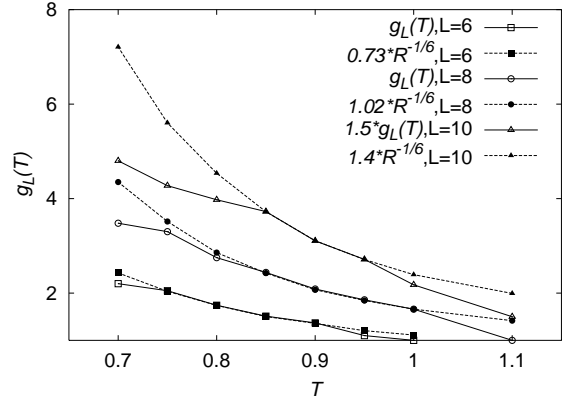}}}
\vskip 0.5cm
\caption{The function $g_{L}(T)$ determined for $I$-$V$ characteristics with
size $L=6$ (open squares), $L=8$ (open circles), $L=10$ (open triangles)
obtained as in Fig.~\ref{fig:scaling}. One notes that the function $g_{L}$ for
these sizes over a limited $T$ interval around KT-transition (at $T \approx
0.90$) is well approximated by $g_{L}\propto R^{-\alpha}$ with $\alpha \approx
1/6$. The data for $L=10$ function $g_{L}(T)$ as well as $A_{L}R^{-\alpha}$ are
multiplied by 1.5 for convenience.}
\label{fig:g_L}
\end{figure}

 The FFH scaling given by Eq.~(\ref{eq:p1}) is correct only in the thermodynamic
limit $\xi/L \rightarrow 0$. However, from a practical point of view there is a
connection between Pierson method, which is based on FFH scaling, 
and the finite-size
scaling introduced by Eq.~(\ref{eq:scale}). If we assume that $\xi$ in
Eq.~(\ref{eq:p1}) is proportional to $R^{-\alpha}$, where $\alpha$ is
$T$-independent constant, the connection between the two different scaling
approach becomes: $g_{L}(T) = A_{L}R^{-\alpha}$ with $A_{L}$ being a constant
which may depend on $L$. In Fig.~\ref{fig:g_L} there are presented three
different functions $g_{L}(T)$ corresponding to $L=6$, 8, and 10. These
functions are determined from the condition that curves corresponding to the
different temperatures should collapse when  plotted as  $E/JR$ vs
$LJg_{L}(T)$.  Since it is well established that the 2D $XY$ model on the
square lattice has the KT transition at $T_{KT} \approx 0.892$
(Ref.~\onlinecite{olsson1}), Fig.~\ref{fig:g_L} shows that $g_{L}(T)$ over a
limited region in the vicinity of the KT transition is very well represented by
the  $R^{-\alpha}$ with $\alpha = 1/6$ for all investigated system sizes.

\begin{figure}
\centering{\resizebox*{!}{5.5cm}{\includegraphics{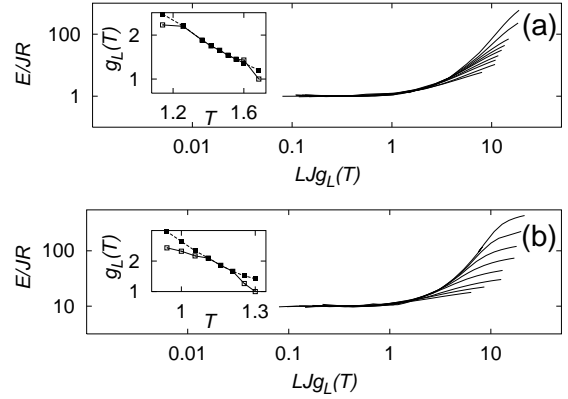}}}
\vskip 0.5cm
\caption{Existence of the scaling in the form Eq.~(\ref{eq:scale}) for modified
2D $XY$ models corresponding to different vortex density. Systems with size
$L=8$ have been simulated with RSJD. The finite-size scaling of the $I$-$V$
characteristics obtained for 2D $XY$ model with the Villain potential in the
temperature range $1.70 \leq T \leq 1.15$ is shown in (a) whereas (b) shows the
same thing for the $p=2$-potential  for $1.30 \leq T \leq
0.95$. The inset on both (a) and (b) shows the function $g_{L}(T)$
determined from the condition of the best data collapse. It is also shown there
that the function $g_{L}$ for both cases over a limited $T$ interval is well
approximated by $g_{L}\propto R^{-\alpha}$ with $\alpha \approx 1/6$.  In
(a)  $g_{L}\propto 0.76 R^{-1/6}$ and in (b)  $g_{L}\propto
1.04 R^{-1/6}$.}
\label{Villain}
\end{figure}

 Next we demonstrate how the finite-size scaling given by Eq.~(\ref{eq:scale})
works for the $I$-$V$ data obtained by simulations of the modified $XY$ model.
These simulations are done with RSJD. Fig.~\ref{Villain}(a) verifies that this
scaling indeed exists for the $XY$ model modified with the Villain type of
potential introduced by Eq.~(\ref{eq:UVillain}). The inset of Fig.~\ref{Villain}(a)
shows the scaling function $g_{L}(T)$ determined by finding the best data
collapse for small values of $LJg_{L}(T)$. One can see that in the vicinity of
$T_{KT}$, which is approximately equals to $1.4$ for this model, $g_{L}(T)$ can
be fitted by $A_{L}R^{-\alpha}$ with $\alpha = 1/6$. The data collapse in
Fig.~\ref{Villain}(b) shows that $E/JR$ is only a function of the scaling
variable $JLg_{L}(T)$ (when $JLg_{L}(T)$ is small enough) for the $XY$ model
modified with $p$-type of potential (Eq.~(\ref{eq:Up})), where $p=2$. The
inset shows the function $g_{L}(T)$ together with $A_{L}R^{-\alpha}$.
Since $T_{KT} \approx 1.15$ for this model one can again see that the scaling
function $g_{L}(T)$ in the vicinity of KT transition is proportional to
$R^{-\alpha}$ with the same exponent $\alpha = 1/6$ as for the original $XY$
and Villain models.

 The crucial difference between the Villain model, the usual $XY$ model and the
$p=2$ $XY$ model in the present context is the vortex density. The
KT-transitions for these models occur at the Coulomb gas temperatures
$T_{c}^{CG} = 0.23$, 0.2, and 0.1, respectively ($T^{CG} = {T \over 2\pi}
\langle U^{\prime\prime}\rangle$ (see Ref.~\onlinecite{minnhagen:rev})). 
Lower $T_{c}^{CG}$ means
higher vortex density. Thus the finite-size scaling property given by
Eq.~(\ref{eq:scale}) appears to be independent of vortex density.
   
\section{Discussion and conclusions} \label{sec:conc}
We have simulated 2D {\it XY} model with three types of dynamics: RSJD, RD, and
MCD. The main conclusion of the paper is  that the qualitative features of the
finite-size scaling given by Eq.~(\ref{eq:scale}) are independent of both the
vortex density and the type of dynamics. Therefore the finite-size scaling
behavior given by Eq.~(\ref{eq:scale}) of the finite-size induced tails of the
$I$-$V$ characteristics appears to be a robust feature. From the comparisons of
the current-voltage characteristics obtained for each type of dynamics we found
that, up to some scale factor, $I$-$V$ curves at a given temperature are
identical over a broad range of external currents. This makes it possible to
use MCD simulations, which are more computer efficient than RSJD and RD
simulations, to obtain $I$-$V$ curves corresponding to both RSJD and RD
dynamics. 

The phase transition for the 2D $XY$-type models are of KT type with
$z=2$. This raises the intriguing question of the origin of the
large $z \approx 6$ obtained by Pierson {\it et al.}~\cite{pierson}. In
Ref.~\onlinecite{medv} it was argued that the Pierson scaling in relation to
the finite-size scaling given by Eq.~(\ref{eq:scale}) corresponds to the
proportionality $g_{L}(T) \sim R^{-\alpha}$ where $1/\alpha$ is the exponent
which corresponds to the ``$z$'' obtained by the Pierson scaling. The reason
for the existence of this scaling like behavior is still unclear.

In the present paper we have shown that, within the class of 2D $XY$-type
models studied, a value $1/\alpha \approx 6$ is obtained independently of the
type of dynamics, as well as, of the vortex density. The origin of this
seemingly robust behavior calls for further investigations. 

\acknowledgments
This work was supported by the Swedish Natural Research Council
through Contract No. FU 04040-332.

\newpage
     
\end{multicols}
\end{document}